High-Throughput Computation of Li-based Battery Material Databases: Chemistry-Processing-Property Relationships

Scott R. Broderick [1,*], Kaito Miyamoto [1,2], and Krishna Rajan [1]

[1]Department of Materials Design and Innovation, University at Buffalo

[2] Toyota Research Institute of North America, Toyota Motor North America, Inc.

*scottbro@buffalo.edu

Abstract

In this paper, we incorporate processing into the data driven property modeling of Li-based spinel battery materials. When considering spinel compounds of the form $LiMe_2O_4$ with Me as a metal or metals, there are 125,000 possible combinations assuming a maximum of three metallic elements. In this work, we focus on capacity for our predicted property, and increase the number of possible combinations to *two million* by incorporating processing into the modeling. Due to the non-linear relationship between processing and property, as well as ensuring proper sensitivity, a new approach which tracks the change with changing processing is introduced. This work provides an invaluable tool for guiding the next generation of battery experiments, providing a significant narrowing of the infinite design space.

## Introduction

A key challenge in machine learning design is the incorporation of processing into the analyses. There have been many successes in the area of data driven chemical design of materials across many material platforms [1-8]. However, due to the complexity of processing, including the consideration of thermodynamic relationships and expanding to a microstructural level, the ability to ‘soft’ model the systems is more challenging. In this paper, we introduce a new approach which allows us to incorporate processing parameters into the training data and to develop high throughput process-chemistry-property relationships.

Our test system here is spinel systems with formula $AB_2O_4$, with Li occupying the A-site, and our primary property of interest here is capacity. The reason for selecting spinel systems is due to their wide usage and beneficial properties in batteries [11-13]. However, the impact of different annealing conditions, for example, have a large impact on the properties of spinels [14-22]. Beyond the development of a descriptor set incorporating a wide series of chemical descriptors, we include annealing time and annealing temperature as additional controllable descriptors or parameters. Therefore, the objective of this work is the prediction of capacity of spinel compounds for use as battery materials, as a function of the metallic elements occupying the B-site, along with time and temperature. As we highlight in this paper, as well as mentioned above, the inclusion of processing parameters is not a trivial issue. We develop new descriptors through a non-linear manifold learning approach, which represent the chemistry/composition of the spinel and the annealing conditions. Through regression applied to these descriptors, we develop a high-throughput model and expanded material property space. This builds on our prior works in data driven design across multiple material systems [23-29], demonstrating the robustness of this logic.

## Method

The specific process developed is shown in Figure 1. The input requires three different pieces of information: two of the inputs include descriptors associated with the controls (ie. chemistry and processing) and which go into graph network analyses, and the other input is the property data which is used for model training and testing. The graph network is built through the IsoMap analysis [30,31]. The details of the IsoMap are provided in the Supplementary Material. The key to this approach is the comparison of the positioning of points within networks as we vary the data input. One network uses only the chemical descriptor set while the other network uses the chemical descriptor set plus time and temperature for processing. These steps provide a parameterization of the data, which is described below. The parameterization then serves as the input into a regression, with the regression providing a high-throughput model. Including additional ‘virtual’ chemistries and processing conditions into the input set results in the development of new parameters, with the model built using the same training data but which is now applicable to the new data and thereby allows us to rapidly predict new property data. The inclusion of ‘virtual’ data results in a minimal decrease of model accuracy, and is therefore still sufficient for the identification of most promising systems or where experiments are needed.

The chemical descriptors are listed in Figure 2. The approach for defining the descriptor values is by scaling the composition of the metal. As we are fixing Li in the A-site and O in the C-site of $AB_2C_4$ spinel stoichiometry, the chemical descriptors only represent the B-site chemistry, with the B-site potentially occupied by multiple elements. The constraint of Li and O will be lifted in future work, as well as the stoichiometry. Further, this approach is assuming random site occupancy of the B-site by the elements, and thus uniform distribution of any lattice strains. This assumption is likely not entirely accurate, but given the purpose of this work is to quickly screen the massive material search space, it is expected that any error introduced by that is within the uncertainty level of the predictions.

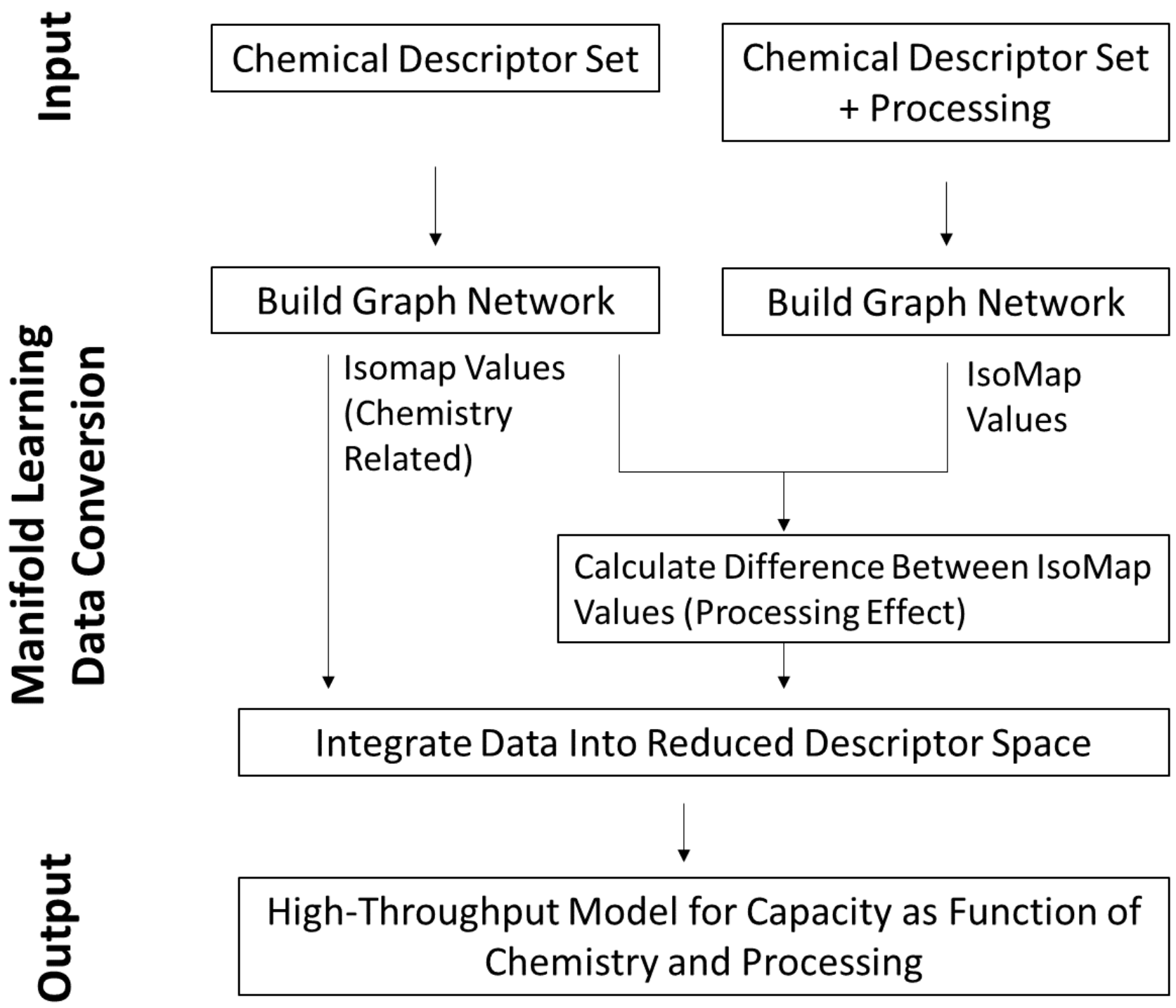

Figure 1. Process for high-throughput prediction of battery material properties. The framework has the additional step of incorporating processing. By having a two level graph network step the impact of processing is accounted for in a non-linear manner, without biasing the model due to the higher dimensionality of chemical descriptors.

The 26D descriptor set corresponding to chemistry constitute the "Chemical Descriptor Set" in Figure 1, while the addition of the processing terms to these constitute the "Chemical Descriptor Set + Processing" matrix. In our case, the first matrix has dimensions of 28 x 26, while the second matrix has dimensions of 68 x 28.

In our case, the search space covers the elements labeled, with the red being elements most represented in the training data while the blue are the least represented, and thus the predictions with them result in higher uncertainty. Also shown is the spinel structure. In this work, we are substituting the blue atoms (B-site in $AB_2X_4$), while the orange atoms (A-site) are always Li and the red atoms (X-site) are always oxygen. The descriptor space described above then constitutes a large dimensionality which is hard to assess in real space. That data is compressed into a graph network, with each point having parameter values which capture non-linear relationships in the real space data. These parameters are then input into a regression which forms the model of property as a linear function of the parameters, although the parameters themselves capture non-linear relationships. As opposed to alternate approaches, such as principal component regression which defines the parameters as principal components, in this case the equation in terms of the initial descriptors is not easily defined.

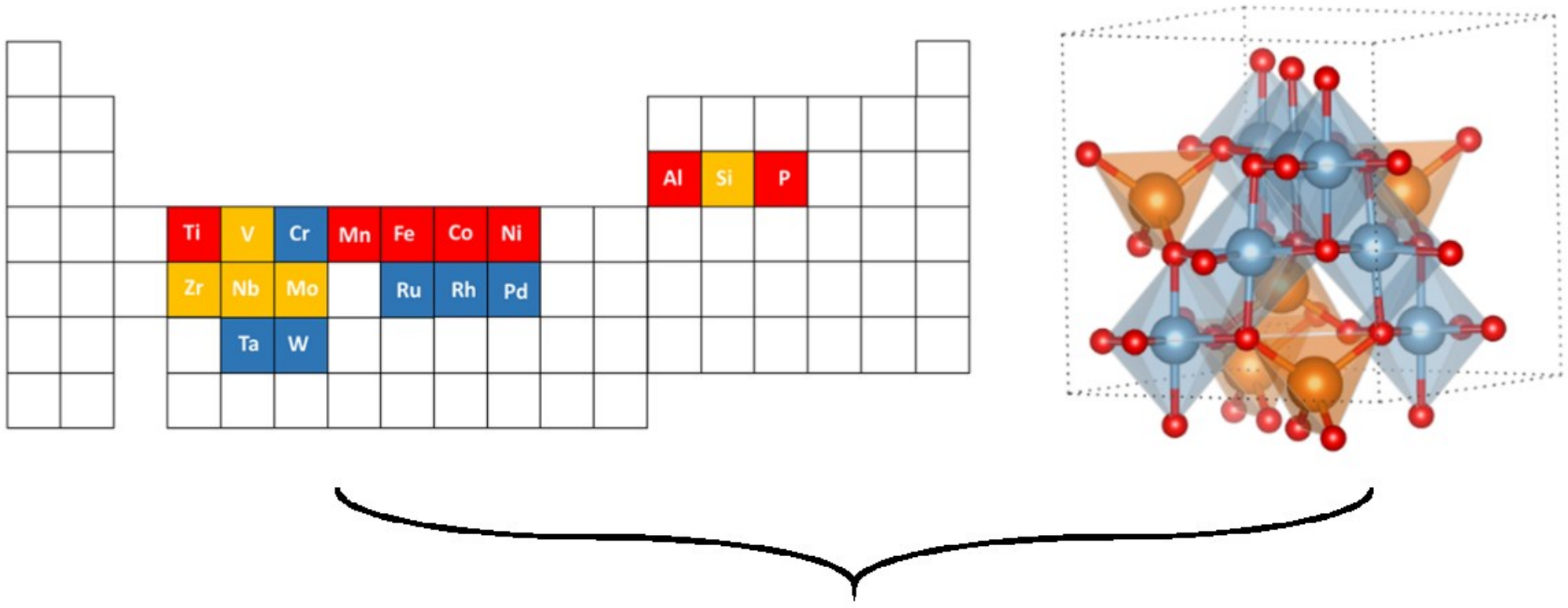

| | | | |
|---|---|---|---|
| 1 | Atomic Number **(AN)** | 14 | Debye Temp (K) **(DT)** |
| 2 | Interatomic distance (Å) **(ID)** | 15 | First Ionization Potential (eV) **(FI)** |
| 3 | Valence electron number, Nv **(VN)** | 16 | Second Ionization Potential (eV) **(SI)** |
| 4 | the *Martynov-Batsanov* electronegativity X $[(eV)^{1/2}]$ **(MBE)** | 17 | Work Function (eV) **(WF)** |
| 5 | Pseudopotential core radii sum $R=r_s+r_p$ **(PR)** | 18 | Atomic Radius (Å) **(AR)** |
| 6 | Covalent Radius (Å) **(CR)** | 19 | Melting Point (K) **(MP)** |
| 7 | Pauling electronegativity **(PE)** | 20 | Boiling Point (K) **(BP)** |
| 8 | Specific Heat (J/gK) **(SH)** | 21 | Density @ 293 K ($g/cm^3$) **(D)** |
| 9 | Heat of Fusion (kJ/mol) **(HF)** | 22 | Molar Volume ($cm^3/mol$) **(MV)** |
| 10 | Heat of Vaporization (kJ/mol) **(HV)** | 23 | Thermal Conductivity at 300K (W/cmK) **(TC)** |
| 11 | Atomic Weight **(AW)** | 24 | Coefficient of Thermal Expansion (10^-6 K^-1) **(CTE)** |
| 12 | Heat Capacity (J/mol.K) **(HC)** | 25 | Bulk Modulus (GPa) **(BM)** |
| 13 | Electrical Conductivity (10^6/cm ohm) **(EC)** | 26 | Shear Modulus ($mu*10^{-6}$), $Kg/cm^2$ **(SM)** |

**+ Annealing Time and Temperature**

Figure 2. Our initial design space. The labeled atoms are what we use in the training data, with the elements occupying the B-sites (shown as the blue atoms in the spinel structure). The 26 listed parameters define the input into our initial analysis (Figure 1), with the added two processing terms used to define processing specific descriptors while accounting for changing chemistry.

## Results

Following the process laid our Figure 1 and using the descriptor set described in Figure 2, we perform a series of IsoMap analyses. The first analysis results in the top network seen in Figure 3. This process follows the standard IsoMap analysis laid out in the Supplementary Material. In this network, each node represents a different chemistry. As processing is not input into this network, we only get one point for each chemistry. The parameters are ordered though in amount of information captured (based on geodesic distances), so that only a few parameters are needed to capture the vast majority of information in the data. In that way, we can minimize the dimensions going in the regression model and thereby minimize the risk of over-fitting the data. The number of nearest neighbors, which is an input into the analysis, is selected such that the addition of another nearest neighbor results in less than 5% additional variance captured, while also ensuring that no short circuiting of the network occurs.

The process is then repeated with processing data included, with the result shown in the bottom network in Figure 3. The difference in position for each chemistry in each network is then used to represent the role of processing in our input data. By defining the descriptors in this way, we account for the role of processing with respect to the chemistry. That is, the change in capacity values is not scaled directly with processing, and therefore the inclusion of processing parameters directly does not represent that relationship. Further, the usage of both sets of parameters helps to add emphasis to the chemistry of the material, which is found to be the most significant characteristic.

To provide an example of calculating the process related descriptors, the case of $LiTi_2O_4$ is highlighted by arrows in Figure 3. In the top network, parameters based solely on the chemistry are defined with the first two parameters visualized in the figure. In the final analysis, three parameters (IsoMap 1 – 3) are used as input for the regression. In the lower figure, another set of parameters are defined which include the chemical descriptors in addition to processing. These parameters from the bottom figure are a function of both chemistry and processing. However, when these parameters alone are used in building the regression model, as discussed below, the accuracy is low, as the model is to weighted towards processing. To address this, we have introduced a new step where we track the change in position due to processing. In this way, the values are on an appropriate scale for processing versus chemistry effect, while still capturing the impact of processing relative to chemical changes.

These processing specific descriptors are then defined as the difference in position between the network with only chemical descriptors and the network with chemical and processing descriptors. The calculation of the processing descriptors for $LiTi_2O_4$ annealed at 600 ºC is shown. For this material, the parameters which are used in the regression model are the first three IsoMap parameters for chemistry only (2.08, -1.63, and the not shown IsoMap3 value of 0.41) and the two processing descriptors calculated here (0.79 and -0.21). These five values are then the new descriptor set for $LiTi_2O_4$ annealed at 600 ºC which we input into the regression analysis to predict capacity. This step is repeated for each node in the lower network of Figure 3.

This process described in Figure 3 corresponds with the "Manifold Learning Data Conversion" step in Figure 1, and results in the "Integrate data into reduced descriptor space" box. A larger number of parameters could be defined and there is no defined rationale in the selection of number of parameters; however, the five parameters described here was found to provide a sufficient combination of accuracy and robustness.

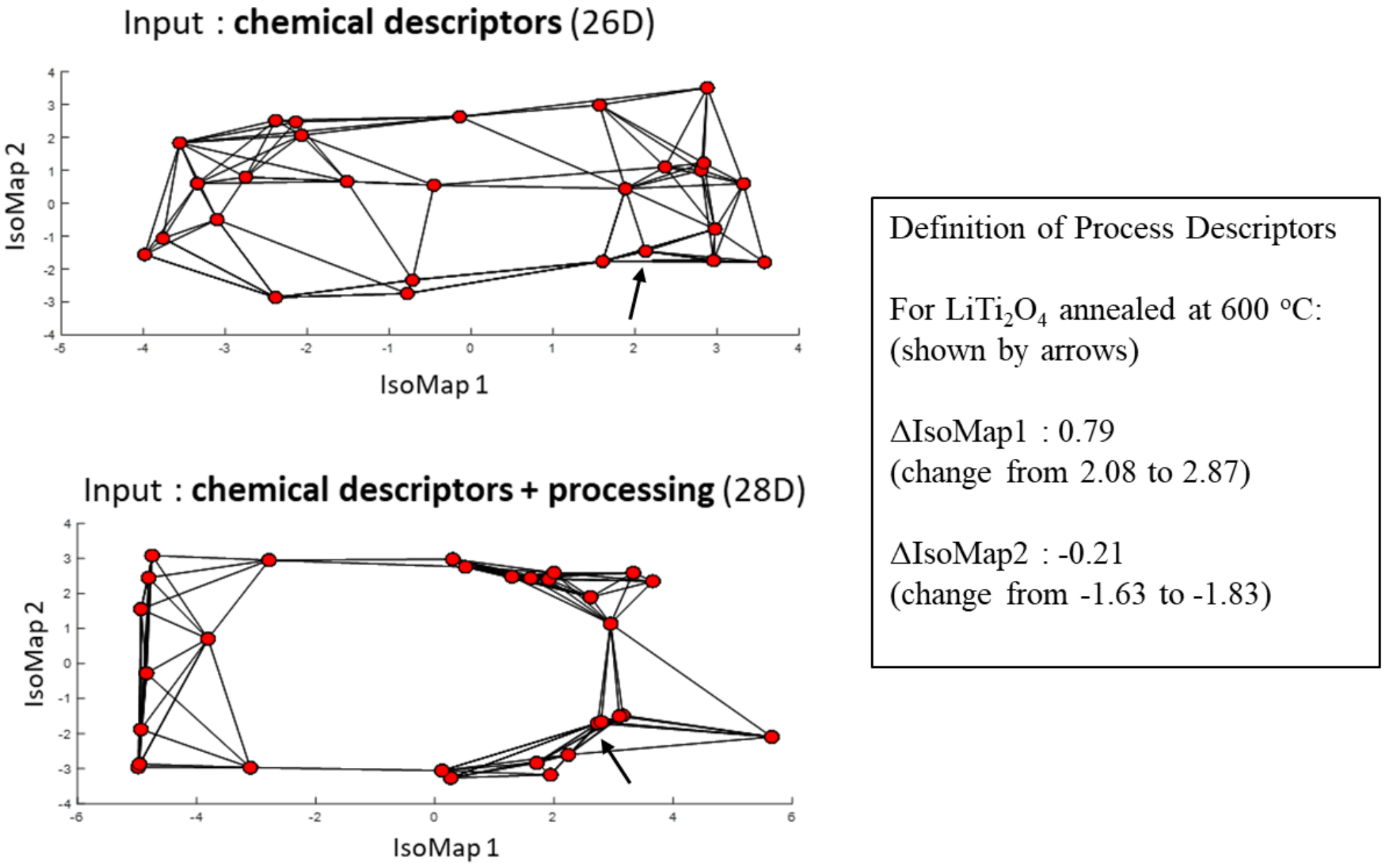

Figure 3. Process for parameterizing the data. The top network is based only on the chemical descriptors, while the lower network uses the same chemical descriptors plus the annealing temperature and time. For purpose of clarity, only the entries corresponding with the top node are shown in the bottom network. The difference in networks due to addition of processing leads to an additional two processing related terms, which are shown for $LiTi_2O_4$ in the inset. These parameters then serve to represent the entire descriptor set with minimal loss of information, and serve as the input into the regression analysis.

The result of the regression model is shown in Figure 4. For ensuring robustness, 20% of the data was kept back for testing the model robustness, with the process repeated 10 times with different cuts of training versus testing data. In all cases, the selection of training versus test data was randomly selected. In all cases, the cross validation $R^2$ value was within 10% of the training value. To further highlight the gain by defining the processing term as we did in Figure 3, the model results for both our approach developed here and the model if we were to use processing as a separate term is shown. The gain in accuracy is significant by following our approach.

In Figure 4, we show the results for two different models. The first model is using the approach laid out in Figure 1 and described in this paper. This results in a model with $R^2$ = 86.6%. To highlight the gain from our approach, we repeated the analysis following a traditional approach (labeled as "Processing as separate term"), where the annealing time and temperature were used as the two additional descriptors for the regression model, in addition to the parameters from the chemistry only network. Thus, time and temperature were used in place of the process descriptors discussed in Figure 3. This resulted in a model with the reduced accuracy of $R^2$ = 69.1%. Therefore, this added step to the framework resulted in a significant accuracy and therefore allows us to reasonably screen the entire chemistry / processing design space rapidly.

We can now predict the capacity value for systems that have not been experimentally measured (ie. 'virtual' materials). We can randomly generate combinations of new chemistries and process conditions, and then repeat the process. While there are no required constraints in defining these compositions; however, in this paper, we limit the 'virtual' systems to those containing the elements in Figure 2. The parameters for 'virtual' systems are developed in the same way with the model linking the parameters and the capacity for only the training systems. From this, we are able to predict the capacity value for millions of compounds, resulting in a huge explosion of data as compared to what is currently available. One point to make is that this work is not meant as a replacement for experiments. Rather, this approach serves to provide massive 'virtual' data which leads to better selecting future experiments, and thus serves to help guide the selection of experiments.

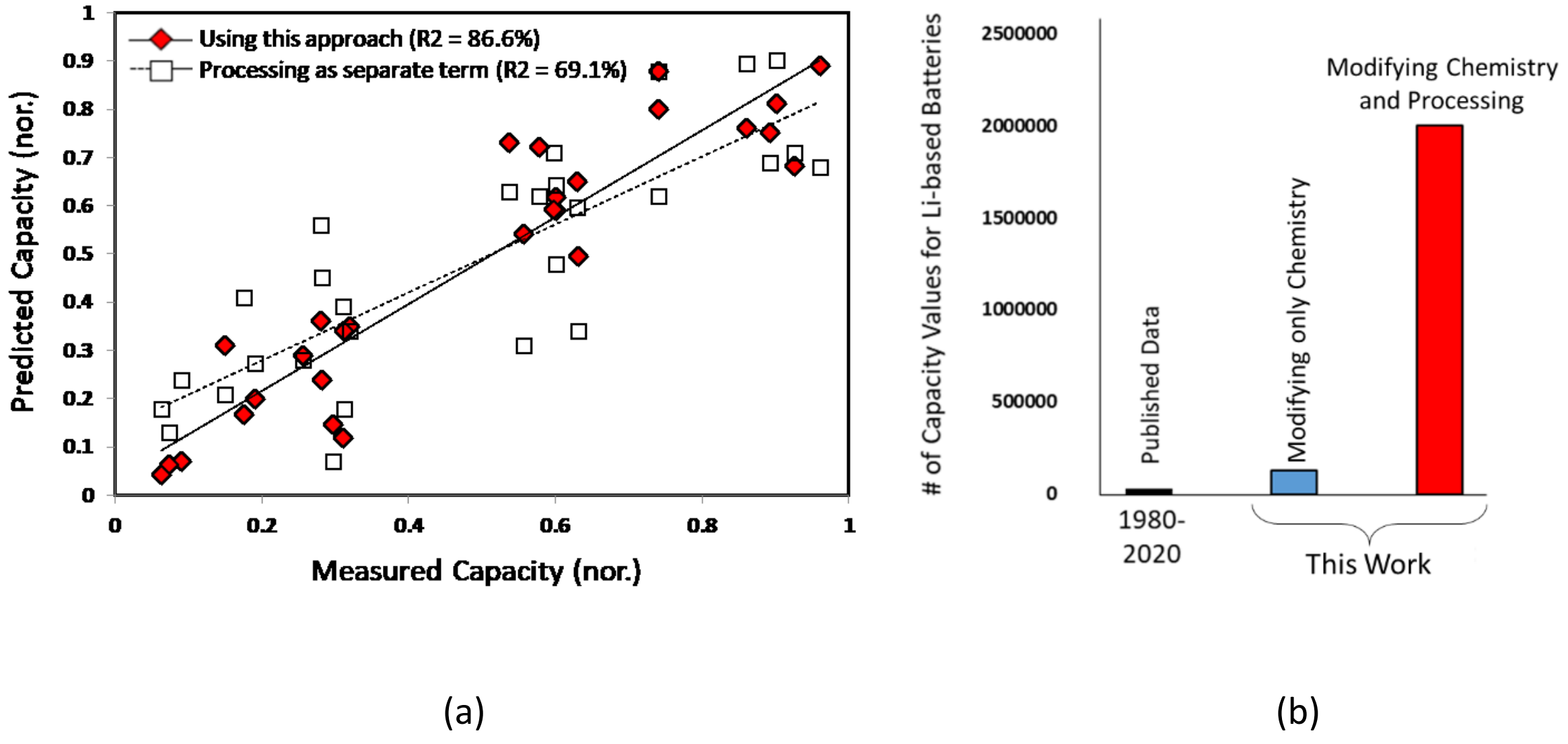

Figure 4. Result from the analysis. (a) Based on this approach, we have a model of high accuracy, which is easily transferrable. By defining the processing descriptors as described in Figure 3, the accuracy of the model has increased from $R^2$ of 69.1% to 86.6%. (b) From this model, we have further significantly expanded the battery material knowledge-base. The number of reported values in literature is relatively small. We have continued that rapid acceleration by incorporating processing terms, thus leading to a larger range of applicable conditions.

Finally, the comparison of models defining processing in different ways is shown in Figure 5. As discussed at the beginning of this paper, the modeling of processing impact on properties is challenging. This figure highlights the challenge by demonstrating how the different approaches to introduce processing result in very different models. We use principal component analysis (PCA) to identify the correlation of the descriptors with capacity. This process for extracting correlations from PCA follows what was described in our prior works [24,32]. This gives us a rough estimate of how much chemistry should impact our model versus processing. The descriptors shown as individual bars here correspond with the list in Figure 1. From this, we identify the chemistry is approximately 50% more important than processing, with the normalized correlation as 0.62 for chemistry and 0.38 for processing. Thus, this defines the weighting of each on our model that is needed. The approach demonstrated in this paper provides a weighting of 0.41 for processing, which closely matches with the expected 0.38. This therefore further justifies the approach developed and highlights that the modeling approach is based on the underlying physics and is not just statistically derived. If we instead consider the "Processing as separate term" model in Figure 4, the role of chemistry in defining the model is only 39%, indicating a loss in describing the underlying physics. Therefore, our work here provides an accurate and physically robust quantitative prediction approach.

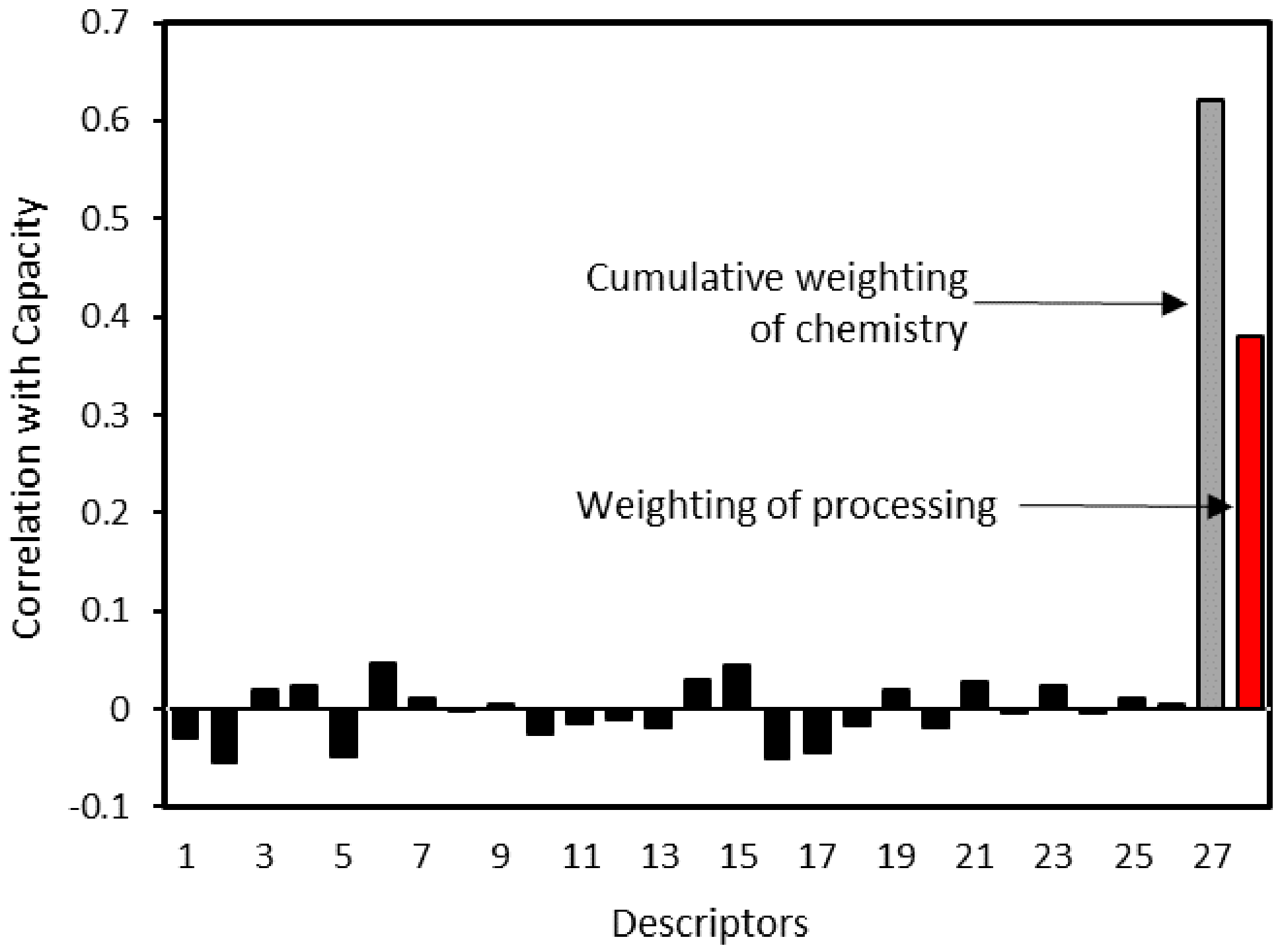

Figure 5. Role of each descriptor, with the descriptor numbers corresponding to Figure 2. The cumulative impact of chemistry on capacity is 62%, as opposed to 38% on processing. The correlations were defined through a Variable Importance Projection analysis [24,32,33].

## Summary

This work introduced a new approach for data driven modeling of battery material data, which also incorporates processing into the prediction. Based on this, we are able to predict capacity for a nearly infinite search space. A screening of the predicted data should serve to help define the next experiments and what the most promising systems and processes are. For the purpose of develop this methodology, numerous constraints have been put in place. These include only looking at capacity, spinel structures, and changing the B-site chemistries. This methodology has been developed in a generic form, and therefore it is expected that future reports will remove these constraints and provide an even larger explosion of battery material data.

## Acknowledgements

We gratefully acknowledge the support of Toyota Motor North America, Inc. SB and KR acknowledge support from the National Science Foundation (NSF) under Grant No. 1640867. KR acknowledges the Erich Bloch Endowed Chair at the University at Buffalo.

Supplementary Material :

## S1. Mathematical Background

The machine learning approach encompasses multiple stages: the development of a relevant descriptor set, the parameterization of data to avoid over-fitting of the model while maintaining the governing physics, and the development of a high-throughput quantitative structure-property relationship (QSAR). In both cases, the properties were not input into the data parameterization, so as to allow for prediction of unknown or 'virtual' materials.

The parameterization of the data was done following a non-linear manifold learning approach, and namely the IsoMap algorithm [1-4]. This approach generates a graph connecting data points on a high dimensional space to their nearest neighbors, mapped out in the high dimensional space, and then fit to a low dimensional manifold. The objective of the Isomap algorithm is to map the distribution of elements in the high dimensional space, represented by the set of data points $\{x_i\} \in R^n$, onto a convex nonlinear manifold $M^d$ of lower dimension $d < n$ and through dimensionality reduction, obtain a two or three dimensional embedding of the elements into a weighted graph. The mapping is carried out such that the geodesic distances between the elements in the higher dimensional manifold is preserved when it is mapped onto the lower dimensional graph, so that the edges of the graph are weighted in their length according to the original geodesic distances. The dissimilarity between alloying elements, which themselves form the vertices of this graph, are captured by these distances between them along the edges that connect them to their nearest neighbors. This mapping can be described in set theory as: $x_i \rightarrow y_i \mid y_i \in M^d,\ d<n,\ s.t.\ \forall(i,j): \left|x_i - x_j\right|_\beta = \left|y_i - y_j\right|_\beta$ where $\beta$ is a norm, representative of the pairwise geodesic distances $d_{ij}$ between any two elements $'i'$ and $'j'$, which is the curvilinear distance along the manifold in $M^d$.

In order to construct the initial graph in $R^n$, we used K nearest-neighbors (KNN), which graphs each point connected by an edge to its $'k'$ nearest neighbors alone $\ni$ $d_{ij} = \infty, \forall \ |i-j| > \mathrm{k}$ . Among semi-supervised learning KNN has been found to perform well compared to other graphs [5] and was, therefore, employed in the present work. In this work the choice of $k$ was optimized by statistically determining the smallest value that could minimize the residual variance | $d_M$ - $d_G$ |, while providing the maximum number of alternative paths. This ensures that the resulting graph is neither over-connected, leading to loss of pairwise geodesic distances, nor are critical neighbors disconnected. For each data point, we also compute the ratio of the distance to its closest and farthest neighbor. The ratios are then averaged over all data points to calculate a scale-invariant, global parameter, Δ, [6] to estimate the measure of uncertainity introduced by sparsity in high dimensional spaces, given that the data points must have sufficient density on the manifold [7]. Δ can range between zero and one and a small value indicates a healthy variance in pairwise distances.

For property prediction, we used the descriptor set and input it into a graph analysis. This provides a set of parameters, which capture the non-linear relationships in the descriptors. The reason for doing this is to reduce the dimensionality of the input without losing information. That is, we use the graph theory for dimensionality reduction, and thereby can perform a regression with limited risk of over-fitting the model. The reason is that we want to create a parameter space which can be used for all systems – that is, we develop a model based on graph theory parameters, and we need these parameters for all systems of interest.

One of the regression approaches applied and reported was using principal component regression (PCR). In PCR, the training data is converted to a data matrix with orthogonalized axes, which are based on capturing the maximum amount of information in fewer dimensions. In this case, as we were predicting for a single property, The relationships discovered in the training data can be applied to a test dataset based on a projection of the data onto a high-dimensional hyperplane within the orthogonalized axis-system. Typical linear regression models do not properly account for the co-linearity between the descriptors, and as a result the isolated impact of each descriptor on the property cannot be accurately known. However, by projecting the data onto a high-dimensional space defined by axes which are comprised of a linear combination of the composite descriptors and also orthogonalized, the impact of the descriptor on the property can be identified independent of all other descriptors.

PCR finds the maximum variance in the predictor variables ($X$) and finds the correlation factors between $X$ and the predicted variables ($Y$) that have maximum variance. The scores of $X$, $t_a$ ($a$=1, 2, …, A=the number of principal components) are calculated as linear combinations of the original variables with the weights $w^*_{ka}$. The multidimensional space of $X$ is reduced to the A-dimensional hyper plane. Since the scores are good predictors of $Y$, the correlation of $Y$ is formed on this hyper plane. The loadings of $X$ ($P$) represent the orientation of each of the components of the hyper plane. Following this, an accurate and high-throughput equation linking the input parameters and the properties are derived.

S2. Descriptor Space

The following (Table S1) provides a list of the input descriptors used in the initial analysis, and provides the list of descriptors from Figure 3. These descriptors have been utilized in our prior work. Of note, these descriptors describe the single element systems, with the values corresponding with the single element systems and in their ground state structures. The scaling of these descriptors to account for multicomponent systems was discussed in the main document. In this case, we limited our search space to $LiMe_2O_4$ and therefore we only scale based on the ratio of elements comprising Me (ie. the B-site in the spinel structure). Future reports will expand to account changing the overall stoichiometry, as well as the A-site and X-site chemistries.

*Table S1. The descriptors used in this analysis. In this paper, we defined descriptors for multi-component systems based solely on the descriptors of the individual elements. In this way, there is no limitation in the 'virtual' design space for which we can apply our model. The uncertainty in models is based solely on the amount of available information in the training or test data, but as this descriptor data is available for nearly all elements, the model can be applied to a nearly infinite chemical space.*

| | | |
|---|---|---|
| Interatomic distance | Pauling electronegativity (EN) | Heat of Vaporization |
| Valence electron number | Martynov-Batsanov EN | Heat Capacity |
| Pseudopotential core radii sum | Melting Point | Melting Point |
| Covalent Radius | Boiling Point | Boiling Point |
| Atomic Radius | Bulk Modulus | Modulus of Elasticity |
| Atomic Weight | Shear Modulus | Electrical Conductivity |
| Molar Volume | Work Function | Thermal Conductivity |
| Density @ 293 K | Specific Heat | Coeff. Thermal Expansion |
| First Ionization Potential | Heat of Fusion | |

## S3. Input Property and Processing Data

*Table S2. The property values input into our analysis. This incorporates chemistry, which we develop the descriptors for using the features in Table S1, as well as processing temperature and time.*

| | capacity (mAh/g) | Processing Temperature (oC) | Processing time (hr) |
|---|---|---|---|
| LiNi0.8Mn0.8Co0.4O4 | 179 | 900 | 12 |
| LiNi0.8Mn0.8Co0.4O4 | 175 | 900 | 12 |
| LiNi0.8Mn0.8Co0.4O4 | 170 | 900 | 12 |
| LiNi0.8Mn0.8Co0.4O4 | 165 | 900 | 12 |
| LiFePO4 | 165 | 500 | 12 |
| LiFePO4 | 160 | 500 | 24 |
| LiFePO4 | 159 | 600 | 12 |
| LiFePO4 | 158 | 600 | 24 |
| LiFePO4 | 150 | 700 | 12 |
| LiFePO4 | 147 | 700 | 24 |
| LiFePO4 | 149 | 800 | 12 |
| LiFePO4 | 127 | 800 | 24 |
| LiFePO4 | 150 | 900 | 12 |
| LiFePO4 | 115 | 900 | 24 |
| LiNiMnO4 | 198 | 600 | 4 |
| LiCo2O4 | 140 | 600 | 6 |
| LiNi0.45Mn1.55O4 | 135 | 600 | 6 |
| LiNi0.45Mn1.55O4 | 130 | 600 | 6 |
| LiNi0.45Mn1.55O4 | 118 | 800 | 6 |
| LiNi0.45Mn1.55O4 | 123 | 800 | 6 |
| LiCu0.05Mn1.95O4 | 118 | 25 | 0 |
| LiCu0.5Mn1.5O4 | 50 | 25 | 0 |
| LiCu0.5Mn1.5O4 | 48 | 25 | 0 |
| LiCu0.5Mn1.5O4 | 46 | 25 | 0 |
| LiCu0.5Mn1.5O4 | 44 | 25 | 0 |
| LiFeSiO4 | 130 | 800 | 12 |
| LiMnSiO4 | 148 | 800 | 12 |
| LiFeSiO4 | 179 | 800 | 12 |
| LiMnSiO4 | 132 | 800 | 12 |
| LiAl0.1Co0.3Ni1.6O4 | 195 | 800 | 12 |
| LiCo2O4 | 184 | 900 | 12 |
| LiMn2O4 | 117 | 900 | 12 |
| LiFePO4 | 170 | 900 | 12 |
| LiTi2O4 | 158 | 600 | 12 |
| LiTi2O4 | 169 | 600 | 24 |
| LiTi2O4 | 172 | 700 | 12 |
| LiTi2O4 | 163 | 800 | 12 |
| LiTi2O4 | 159 | 800 | 6 |
| LiTi2O4 | 155 | 900 | 6 |
| LiTi2O4 | 161 | 900 | 12 |
| LiTi2O4 | 146 | 900 | 24 |
| LiTi2O4 | 151 | 900 | 18 |
| LiNi0.5Mn1.5O4 | 128 | 1000 | 24 |
| LiNi0.5Mn1.5O4 | 82 | 1000 | 24 |
| LiMn1.5Ni5O4 | 131 | 900 | 12 |
| LiMn1.5Ni0.42Fe0.08O4 | 132 | 900 | 12 |
| LiMn1.42Ni0.42Fe0.16O4 | 127 | 900 | 12 |
| LiMn1.5Ni0.34Fe0.16O4 | 124.5 | 900 | 12 |
| LiNi0.5Mn1.5O4 | 122 | 900 | 24 |
| LiNi0.45Cr0.05Mn1.5O4 | 119 | 900 | 24 |
| LiNi0.5Mn1.5O4 | 112 | 900 | 24 |
| LiNi0.5Mn1.5O4 | 108 | 900 | 24 |
| LiNi0.5Mn1.5O4 | 111 | 900 | 18 |
| LiNi0.5Mn1.5O4 | 128 | 900 | 12 |
| LiNi0.45Cr0.05Mn1.5O4 | 122 | 900 | 6 |
| LiTi2O4 | 162 | 25 | 0 |
| LiMnTiO4 | 151 | 25 | 0 |
| LiMn2O4 | 148 | 25 | 0 |
| LiCo2O4 | 147 | 25 | 0 |
| LiCoMnO4 | 146 | 25 | 0 |
| LiFe2O4 | 142 | 25 | 0 |
| LiTiVO4 | 141 | 25 | 0 |
| LiFeNiO4 | 132 | 25 | 0 |
| LiFeTiO4 | 131 | 25 | 0 |
| LiCoNiO4 | 129 | 25 | 0 |
| LiNi0.5Mn1.5O4 | 122 | 25 | 0 |
| LiNi2O4 | 117 | 25 | 0 |
| LiCu2O4 | 115 | 25 | 0 |
| LiCu0.5Mn1.5O4 | 105 | 25 | 0 |